Item Parameter Recovery for the Two-Parameter Testlet Model with Different

Estimation Methods

*Yong Luo*

*National Center for Assessment in Saudi Arabia*

## Abstract

The testlet model is a popular statistical approach widely used by researchers and practitioners to address local item dependence (LID), a violation of the local independence assumption in item response theory (IRT) which can cause various deleterious psychometric consequences. Same as other psychometric models, the utility of the testlet model relies heavily on accurate estimation of its model parameters. The two-parameter logistic (2PL) testlet model has only been systematically investigated in the psychometric literature regarding its model parameter recovery with one full information estimation methods, namely Markov chain Monte Carlo (MCMC) method, although there are other estimation methods available such as marginal maximum likelihood estimation (MMLE) and limited information estimation methods.

In the current study, a comprehensive simulation study was conducted to investigate how MCMC, MMLE, and one limited information estimation method (WLSMV), all implemented in Mplus, recovered the item parameters and the testlet variance parameter of the 2PL testlet model. The manipulated factors were sample size and testlet effect magnitude, and parameter recovery were evaluated with bias, standard error, and root mean square error. We found that there were no statistically significant differences regarding parameter recovery between the three methods. When both sample size and magnitude of testlet variance were small, both WLSMV and MCMC had convergence issues, which did not occur to MCMC regardless of sample size and testlet variance. A real dataset from a high-stakes test was used to demonstrate the estimation of the 2PL testlet model with the three estimation methods.

## Introduction

As currently the dominant measurement framework used in large-scale educational testing, item response theory (IRT; Lord, 1980) boasts over the classical test theory (CTT) theoretical advantages such as parameter invariance (Hambleton & Swaminathan, 1985). For such advantages to materialize, assumptions of IRT need to be satisfied. One pivotal assumption of item response theory (IRT) is local item independence (LII), which stipulates that an examinee's responses to any pair of items should be independent after conditioning on the latent variable of interest. The assumption of LII, however, is often too stringent to be tenable in practical situations.

One common situation in the context of educational testing where LII assumption may be violated is with items sharing a common stimulus such as a reading comprehension passage or a mathematical diagram. Take a reading comprehension test as an example, it is found that items belonging to the same passage tend to be correlated with each other even after conditioning on the latent construct, a phenomenon known as local item dependence (LID; Yen, 1994). LID results in violation of LII, and failure to address LID can lead to serious psychometric consequences such as biased estimation of item parameters (Ackerman, 1987), overestimation of test reliability (Sireci, Thissen, & Wainer, 1991), equating bias (Tao & Cao, 2016), and erroneous classification of examinees into incorrect performance categories (Zhang, 2010). As the psychometric consequences of LID are grave, various methods have been proposed to detect and address LID (e.g., Bradlow, Wainer, & Wang, 1999; Braeken, Tuerlinckx, & De Boeck, 2007; Hoskens & De Boeck, 1997; Ip, 2002; Wilson & Adams, 1995). Among them, the testlet model (Wainer, Bradlow, & Wang, 2007) is probably one of the most popular approaches. Due to its popularity and wide usage in the psychometric community, it is important to evaluate whether the model parameters of a testlet model can be accurately estimated. In addition, when several estimation methods are available (which is the case for the testlet model), it is of methodological interest to compare which method leads to the most accurate estimation and provide researchers and practitioners with advice regarding which estimation method should be used under a given data condition. The current paper is intended to be such a comparison study that compares the performances of three estimation methods (which will be discussed in the following) for the testlet model.

So far the testlet model has been mainly estimated with Markov chain Monte Carlo (MCMC) algorithm (e.g., Koziol, 2016; Li, Bolt, & Fu, 2006) and marginal maximum likelihood estimation (MMLE; Bock & Aitkin, 1981) method (e.g., Jiao, Wang, & He, 2013; Li, Li, & Wang, 2010), both of which are full-information estimation methods. Limited-information methods which are commonly used in categorical confirmatory factor analysis (CCFA; Wirth & Edwards, 2007) can also be readily employed for estimation of IRT models due to the well-known mathematical equivalence between CCFA and IRT models that do not include the pseudo-guessing parameter (Kamata & Bauer, 2008; Takane & de Leeuw, 1987). Studies (e.g., Bolt, 2005; Knol & Berger, 1991) have shown that for multidimensional IRT (MIRT; Reckase, 2009) model estimation, limited-information methods such as the unweighted least square (ULS) method (McDonald, 1982) implemented in NOHARM (Fraser, 1988) and the generalized least square (GLS) method implemented in both Mplus (Muthén & Muthén, 1998-2012) and LISREL (Jöreskog & Sörbom, 1993) performed as well as, if not better than, the full-information maximum likelihood method (Bock, Gibbons, & Muraki, 1988) implemented in TESTFACT (Wilson, Wood, & Gibbons, 1998). As will be shown later, since the testlet model is a special case of the bi-factor model, which is a member of the MIRT family, it is expected that limited-information estimation methods should at least estimate the testlet model as accurately as full-information methods do. In this article, we focus on one such a limited information method, namely weighted least square adjusted by mean and variance (WLSMV; Muthén, du Toit, & Spisic, 1997), the default estimation method for categorical data in Mplus that has been shown to perform better than other limited information methods (e.g., Beauducel & Herzberg, 2006; Flora & Curran, 2004; Yang-Wallentin, Jöreskog, & Luo, 2010). It should be noted that Mplus provides several limited information estimation methods other than WLSMV (see, for example, Paek, Cui, Gubes, & Yang, 2017).

While the three-parameter logistic (3PL) MIRT model is more generalized than the 2PL MIRT model, the latter is often preferred in practice. Such preferences can be attributed to the fact that the pseudo-guessing parameter in the 3PL MIRT models is usually difficult to estimate, and researchers (e.g., McLeod, Swygert, & Thissen, 2001; Stone & Yeh, 2006) often resort to a two-step procedure, in which the pseudo-guessing parameter is first estimated with a unidimensional 3PL model and then plugged into a MIRT software program for the estimation of the remaining model parameters. Such a two-step estimation procedure, however, has been shown to be inferior to simultaneous estimation of all parameters in the 3PL MIRT model through MCMC estimation (Zhang & Stone, 2004). As a special case of MIRT, the 3PL testlet model also faces such issues. Although with MCMC algorithm all the model parameters can be estimated simultaneously, in practice researchers and practitioners sometimes face model convergence issue with the 3PL testlet model estimated with MCMC algorithm and resort to the 2PL testlet model (e.g., Eckes, 2014). As stated by Wainer, Bradlow, and Wang (2007, p. 136), the estimation difficulty of the 3PL testlet model is mainly due to "weak identifiability, that is, presence of multiple different triplets of for which the ICCs are close". Consequently, they recommend that the 3PL testlet model should be used cautiously.

To the best of our knowledge, there is only one simulation study in the psychometric literature that systematically compares different estimation methods for a testlet model: Jiao, Wang, and He (2013) compared MCMC, MMLE, and the sixth-order Laplace approximation (Raudenbush, Yang, & Yosef, 2000) as estimation methods for the one-parameter logistic (1PL) testlet model. However, they did not investigate the performance of any limited information method and consequently, how limited information methods perform regarding the testlet model parameter recovery remains unknown. In addition, the 1PL testlet model is more stringent than the 2PL testlet model in that the item discrimination parameter is constrained to be the same across items.

In this article, we compare the performances of MCMC, MMLE, and WLSMV as estimation methods of the 2PL testlet model. Among the three methods, MCMC (the BAYES estimator) and MMLE (the MLR estimator) implemented in Mplus are both full information method in that they use the full multi-way frequency table (or in the IRT terminology, the whole response vector), and WLSMV in Mplus (the WLSMV estimator) is a limited information method in that it uses only the two-way frequency table. It should be noted that MCMC itself does not necessarily entail a full information method (for an illustration of using MCMC as a limited information method, see, for example, Bolt, 2005). In addition, it is worth noting that when a MMLE estimator is used, Mplus employs the full-information bi-factor analysis method (Gibbons & Hedecker, 1992) to simplify the multidimensional integration issue to two-dimensional integration and hence reduce the computation burden.

The remainder of the articles is structured as follows. First, we demonstrate that 2PL testlet model is a special case of the bi-factor model. Second, we provide a literature review on (a), simulation studies focusing on parameter recovery of IRT models with WLSMV, and (b), simulation studies that investigate parameter estimation of the testlet model. Third, we present a simulation study conducted to compare the item parameter recovery of the 2PL testlet model with the three estimation methods. Fourth, we use a real data set to demonstrate the estimation of the 2PL testlet model with these methods. Finally, we conclude the current paper with discussions and recommendations regarding the parameter estimation of the 2PL testlet model in practice.

## 2PL Testlet Model as a Special Case of the Bi-factor Model

For the 2PL testlet model, it takes the form

$$p_j(\theta_i) = \frac{1}{1+e^{-a_j(\theta_i-b_j-\gamma_{id(j)})}}, \quad (1)$$

where $p_j(\theta_i)$ is the probability of person $i$ giving a correct response to item $j$, $\theta_i$ is person $i$'s latent ability, $\gamma_{id(j)}$ is person $i$'s latent ability on testlet $d$, and $a_j$, and $b_j$ are the discrimination and difficulty parameters of item $j$. For $\gamma_{id(j)}$, its variance $\sigma^2_{\gamma_{jd(i)}}$ indicates the magnitude of testlet effect, or LID among items within the same testlet.

The bi-factor model (Gibbons & Hedeker, 1992) is given as

$$p_j(\theta_i) = \frac{1}{1 + e^{-(a_{jg}\theta_g - b_j + a_{js}\theta_s)}} \ , \tag{2}$$

where $a_{jg}$ and $a_{js}$ are the item discrimination parameters on the general and specific factors for item $j$, $b_j$ is the intercept parameter for item $j$, $\theta_{jg}$ and $\theta_{js}$ are examinee $i$'s latent abilities on the general and specific factors for item $j$.

The 2PL testlet model in equation 1 can be viewed as a special case of the more general bi-factor model in equation 2 in that the discrimination parameter on a secondary factor $a_{js}$, which is freely estimated in equation 2, is constrained to be the product of $\sigma_{\gamma_{id(j)}}$ and $a_j$ in equation 1 (e.g, Li, Bolt, & Fu, 2006; Rijmen, 2010). Consequently, the 2PL testlet model can be viewed as a constrained bi-factor model in the framework of CCFA that can be estimated with limited-information methods.

Studies Using WLSMV for IRT Estimation

WLSMV is a robust version of the WLS method (Muthen, 1984) that is commonly used for CCFA parameter estimation. In contrast to large sample sizes usually required by WLS, WLSMV has been shown to work well with small samples in the context of CCFA (e.g., Flora & Curran, 2004). In the following we review two simulation studies that investigate the performances of WLSMV as an IRT estimation method.

Finch (2010) conducted a comprehensive simulation study to compare how well the ULS method in NOHARM and the WLSMV estimator in Mplus recovered the item parameters of two-dimensional MIRT models with simple structures. The manipulated factors included number of items (15, 30, 60), number of examinees (250, 500, 1000, 2000), distribution of latent trait (normal, skewed), inter-factor correlation (0, 0.3, 0.5, 0.8), and pseudo-guessing (absent, present), and 1,000 replications were conducted within each condition. He found that when pseudo-guessing was present in the generating model, WLSMV recovered the item discrimination and difficulty parameters less well than ULS, because Mplus does not allow users to provide pseudo-guessing values while NOHARM does. When the generating model was the 2PL model, WLSMV and ULS performed comparably. In addition, he found that the performances of both WLSMV and ULS were susceptible to the distribution of latent trait, and with skewed latent trait distribution, the item parameter recovery quality worsened. With the increase of sample size, the accuracy in item discrimination parameter recovery improved for both WLSMV and ULS.

Paek et al. (2017) conducted another simulation study to investigate the performances of nine estimators in Mplus (ML-Logit, ML-Probit, MLR-Logit, MLR-Probit, MLF-Logit, MLF-Probit, WLS-Probit-Theta, WLSMV-Probit-Theta, and ULS-Theta) regarding parameter estimation of the 2PL IRT model. These nine estimators can be viewed as the combination of different use of information (full-information vs limited-information), different link function (logit vs probit), and different parameterization in the limited-information case (Theta vs Delta). They manipulated the number of test items (11, 22, 44) and sample size (200, 500, 1000), which resulted in a fully-crossed design with nine simulation conditions. Within each condition 4,000 replications were conducted. They found that WLS-Probit-Theta had serious convergence issues and performed the worst except when the number of items was small and the sample size was large. ML-Probit, MLR-Probit, MLF-Probit, WLSMV-Probit-Theta, and ULS-Theta outperformed the other three estimators regarding the estimation of the item discrimination parameter. When the sample size was 500 or greater, the five methods produced highly similar results; with a sample size of 200 and a test length of 11 or 22 items, WLSMV-Probit-Theta had slightly worse standard error estimation than the other four methods.

Studies on Testlet Model Estimation

Bradlow, Wainer, and Wang (1999) conducted a small-scale simulation to check whether their proposed 2PL testlet model could be accurately estimated with MCMC algorithm. They simulated a test of 60 items (30 independent items and 30 testlet items) taken by 1000 examinees, and among the 30 testlet items, manipulated the number of items per testlet (5 and 10) and the testlet effect (0.5, 1, and 2) to result in a 2 by 3 simulation study. They found that the item and ability parameters were accurately recovered by the Gibbs sampler. Probably because the computational burden was prohibitive for the computing power back then, they only generated one dataset in each simulation condition.

In a follow-up study, Wang, Bradlow, and Wainer (2002) extended the 2PL testlet model to a more general Bayesian testlet model that can accommodate mixed format tests containing both dichotomous and polytomous items. They conducted a simulation study with nine conditions to investigate whether the general Bayesian testlet model could be accurately estimated. The number of examinees was fixed at 1000 and the number of items was fixed at 30 with 12 independent binary items and 18 testlet items. Among the 18 items, they manipulated the number of items per testlet (3, 6, and 9), the number of response categories (2, 5, and 10), and the testlet effect (0, 0.5, and 1). It should be noted that they did not use a fully-crossed design, which would result in 27 simulation conditions, but instead adopted a Latin Square design that reduced the number of simulation conditions to nine. They ran five replications within each condition. They found that their general Bayesian testlet model could be accurately estimated with the Gibbs sampler.

Jiao, Wang, and He (2013) compared three estimation methods for the 1PL testlet model, namely MCMC implemented in WinBUGS (Lunn, Thomas, Best, & Spiegelhalter, 2000), MMLE implemented in ConQuest (Wu, Adams, & Wilson, 1998), and the sixth-order Laplace approximation implemented in HLM6 (Raudenbush, Bryk, Cheong, & Congdon, 2004), in a simulation study consisting of three simulation conditions. With the number of examinees fixed at 1000 and number of items fixed at 54 (nine items within in each of the six testlets), they manipulated the variance of testlet effect (0.25, 0.5625, and 1) and generated 25 datasets within each simulation condition. Their found that choice of the estimation method significantly affected parameter recoveries. Specifically, the Laplace method resulted in the smallest bias of testlet variance estimation and the smallest random error of ability parameter estimation; the MMLE method resulted in the smallest bias of ability variance estimation; the MCMC method resulted in the smallest bias of item difficulty estimation.

Wang and Wilson (2005) conducted a comprehensive simulation study to investigate whether the Rasch testlet model could be accurately estimated with MMLE implemented in ConQuest. The four factors they manipulated include item type (dichotomous, polytomous, and a mixture of both), number of items within each testlet (5 and 10 for dichotomous items; 3 and 6 for polytomous items), sample size (200 and 500), and testlet effect (0.25, 0.50, 0.75, and 1). The number of items varied with the item type: when the item type was dichotomous, the number of items was 40; when it was polytomous, the number of items was 24; with mixed item format, the number of items was 36 (24 dichotomous and 12 polytomous items). One hundred replications were conducted under each simulation condition. They found that all parameters in the Rasch testlet model could be accurately estimated with MMLE and, with the increase of sample size, the estimation accuracy improved correspondingly.

To sum up, while there are studies that either focus on one specific estimation method or compare two different estimation methods (MCMC vs MMLE) for the 1PL testlet model, neither are there studies that investigate the performances of limited-information methods for parameter estimation of any testlet model, nor are there any comparison studies that compare the performances of MCMC and MMLE for the 2PL testlet model.

## Methods

### Simulation Design

We conducted a Monte Carlo simulation study to compare how well MCMC, MMLE, and WLSMV recovered the item parameters in a 2PL testlet model. The number of items was fixed to 30, and manipulated factors include sample size (SS; 500, 1000, 2000) and testlet variance (TV; 0.25, 0.5, 1), which resulted in a fully-crossed design with nine simulation conditions. We set the number of items within each testlet to five to mimic the typical number of items nested within a reading comprehension passage, thus resulting in six testlets in total for each dataset.

### Data Generation

Similar to Bradlow, Wainer, and Wang (1999), item discrimination and difficulty parameters were generated from a normal distributions $N(1, 0.2)$ and a standard norm distribution $N(0, 1)$, respectively. Table 1 lists the generated item parameters that were used across all simulation conditions. The latent trait parameters were generated from a standard normal distribution $N(0, 1)$, and the person specific testlet effect was generated from a normal distribution with a mean of zero and variance equal to one of the three possible testlet variance values determined by the specific simulation condition. It should be noted that for the same sample size, the same set of latent ability parameter values was used for data generation. One hundred datasets were generated based on equation 1 for each simulation condition.

Table 1

*Item Parameter Used for Data Generation*

| Item | a | b | Item | a | b |
|------|------|-------|------|------|-------|
| 1 | 1.17 | -1.55 | 16 | 0.76 | 0.35 |
| 2 | 0.61 | -1.29 | 17 | 0.98 | -1.24 |
| 3 | 0.67 | 1.44 | 18 | 0.70 | 1.30 |
| 4 | 1.16 | 1.86 | 19 | 0.90 | 0.83 |
| 5 | 1.06 | -0.90 | 20 | 0.58 | 0.06 |
| 6 | 0.69 | 0.05 | 21 | 0.66 | -0.41 |
| 7 | 0.81 | -0.88 | 22 | 0.84 | 1.09 |
| 8 | 0.95 | -0.62 | 23 | 0.81 | 0.01 |
| 9 | 0.51 | 1.89 | 24 | 0.77 | -1.06 |
| 10 | 0.88 | 0.09 | 25 | 0.50 | 0.89 |
| 11 | 0.78 | 0.20 | 26 | 0.97 | 0.62 |
| 12 | 0.96 | -0.19 | 27 | 0.62 | -0.17 |
| 13 | 1.21 | 1.89 | 28 | 0.70 | -0.81 |
| 14 | 0.90 | -0.50 | 29 | 0.82 | -0.12 |
| 15 | 0.94 | 0.27 | 30 | 0.77 | -0.43 |

Estimation Procedure

Mplus was used for estimation with its Bayes, MLR, and WLSMV estimators, which

correspond to MCMC, MMLE, and WLSMV estimation methods, respectively. For the MLR

estimator with logit as the default link function in Mplus, we used probit as the link function to

be consistent with WLSMV and Bayes. It should also be noted that when Bayes was specified as

the estimator, the default priors implemented in Mplus were used: a normal distribution $N(0,5)$

for both the factor loading and item threshold parameters, and an inverse gamma distribution

$IG(-1,0)$ for the testlet variance parameter. As indicated by Asparouhov and Muthén (2010), $IG(-1,0)$ is equivalent to a uniform distribution $unif(0, \infty)$ and hence uninformative. We specified

Mplus to run a minimum of 20,000 iterations for each of four chains, which means that if the model did not converge after 20,000 iterations, Mplus would keep running the four chains until convergence. It should be noted that Mplus employs a convergence diagnostic index called the potential scale reduction factor (PSRF; Gelman & Rubin, 1992) to assess model convergence.

As Mplus does not produce IRT-based parameter estimates for the 2PL testlet model, we used the same procedures as demonstrated by Luo (2017) to convert the factor loading and item threshold estimates in Mplus outputs to the corresponding IRT model parameters. Specifically, for the WLSMV estimator, the following two equations (McDonald, 1999) were used to convert the estimated parameters in Mplus output:

$$a_j = \frac{1.702\lambda_{jWLSMV}}{\sqrt{1 - \lambda_{jWLSMV}'\varphi\lambda_{jWLSMV}}} , \tag{3}$$

$$b_j = \frac{-1.702\tau_{jWLSMV}}{a_j\sqrt{1 - \lambda_{jWLSMV}'\varphi\lambda_{jWLSMV}}} , \tag{4}$$

where $a_j$ and $b_j$ are the discrimination and difficulty parameters of item $j$, $\lambda_{jWLSMV}$ and $\tau_{jWLSMV}$ are the vector of factor loading parameters and item threshold parameter estimated with the WLSMV estimator, and $\varphi$ is the factor covariance matrix. The constant 1.702 is used to convert item parameters from the normal metric to the logistic metric.

For the MLR and Bayes estimators, before equations 2-3 can be used for parameter conversion, the estimates of factor loadings $\lambda_{jBayes/MLR}$ and item threshold $\tau_{jBayes/MLR}$ need to be converted to $\lambda_{jWLSMV}$ and $\tau_{jWLSMV}$ using $R_j^2$, which is the proportion of variance in item $j$ that is accounted for by the latent factor:

$$\lambda_{jWLSMV} = \lambda_{jBayes/MLR}\sqrt{1 - R_j^2} , \tag{5}$$

$$\tau_{jWLSMV} = \tau_{jBayes/MLR} \sqrt{1 - R_j^2} \ . \tag{6}$$

Outcome Variables

The accuracy of parameter recovery was evaluated in terms of Bias, standard error (SE), and root mean square error (RMSE) regarding the item parameters and the testlet variance parameter. Bias, SE, and RMSE are indicative of systematic error, random error, and total error of estimation, respectively. They are defined as

$$Bias(\hat{\pi}) = \frac{\sum_1^R (\hat{\pi}_r - \pi)}{R}, \tag{7}$$

$$SE(\hat{\pi}) = \sqrt{\frac{\sum_1^R (\hat{\pi}_r - \bar{\hat{\pi}})^2}{R}}, \tag{8}$$

and

$$RMSE(\hat{\pi}) = \sqrt{\frac{\sum_1^R (\hat{\pi}_r - \pi)^2}{R}}, \tag{9}$$

where $\pi$ is the true model parameter, $\hat{\pi}_r$ is the estimated model parameter for the $r$th replication, $\bar{\hat{\pi}}$ is the mean of model parameter estimates across replications, and $R$ is the number of replications.

## Results

Model Convergence

As the Bayes estimator had no convergence issue regardless of the sample size and the magnitude of testlet variance, in the following discussion we focus on the WLSMV and MLR estimators. For the WLSMV estimator in Mplus, the occurrence of Heywood case (occurrence of negative residual variance estimates) is indicative of model non-convergence; for the MLR estimator, Mplus generates a warning message stating that the model did not converge. It should be noted that when Heywood case occurs with WLSMV, Mplus still produces parameter estimates that cannot be trusted; when model does not converge with MLR, Mplus does not produce any estimates at all. For all the analyses regarding model parameter recovery in the next section, only those datasets without estimation convergence issues were used for the computation of Bias, SE, and RMSE.

With small sample size (SS = 500) and small testlet variance (TV = 0.25), in 26 replications the WLSMV estimator encountered Heywood case, and in 10 replications the MLR estimator failed to converge; with the same sample size but medium testlet variance (TV = 0.50), the number of times Heywood case occurred with the WLSMV estimator decreased to 10 times, while model non-convergence did not occur at all for the MLR estimator. With medium sample size (SS=1000) and small testlet variance, in 10 replications Heywood case occurred with the WLSMV estimator, and the MLR estimator had no model non-convergence issue. With large sample size (SS = 2000), neither WLSMV nor MLR encountered model non-convergence regardless of the magnitude of testlet variance. It seems that for the 2PL testlet model, both WLSMV and MLR tend to have model convergence issues with small sample size and small testlet variance, but the increase of either sample size or testlet variance tends to reduce the occurrence of model non-convergence for both estimators.

Testlet Variance Recovery

Table 2 lists descriptive statistics of estimation biases of the testlet variance parameter. To evaluate the effect of different manipulated factors on estimation biases, an analysis of variance with Bias as the dependent variable was conducted. The results indicated that the main effect of sample size was statistically significant, $F(2,135) = 9.470$, $p < 0.001$, $f = 0.375$. Tukey *post hoc* tests revealed that estimation biases of testlet variance with SS = 500 were significantly greater than those with SS = 1000 ($p = 0.001$) and SS = 2000 ($p = 0.001$). There was no significant difference in estimation biases between SS = 1000 and SS = 2000.

Another statistically significant main effect was the magnitude of testlet variance, $F(2,135) = 8.172$, $p < 0.001$, $f = 0.348$. Tukey *post hoc* tests revealed that estimation biases of testlet variance when testlet variances were generated to be large were significantly greater than those with medium ($p = 0.001$) and small testlet variance ($p = 0.003$). There was no significant difference in estimation biases between medium and small testlet variance.

The interaction between sample size and magnitude of testlet variance was also statistically significant, $F(4,135) = 14.600$, $p < 0.001$, $f = 0.658$. The main effect of estimation method was not statistically significant, $F(2,135) = 2.576$, $p = 0.080$.

Table 2

*Bias in Testlet Variance Estimation*

| SS | TV | WLSMV | | MLR | | BAYES | |
|----|----|-------|-----|-----|-----|-------|-----|
| | | Mean | SD | Mean | SD | Mean | SD |
| | L | 0.0071 | 0.0042 | 0.0092 | 0.0042 | 0.0104 | 0.0039 |
| 500 | M | -0.0021 | 0.0025 | -0.0014 | 0.0022 | 0.0005 | 0.0021 |
| | S | 0.0008 | 0.0064 | -0.0008 | 0.0068 | 0.0026 | 0.0057 |
| | L | -0.0007 | 0.0047 | 0.0006 | 0.0047 | 0.0011 | 0.0047 |
| 1000 | M | -0.0004 | 0.0016 | 0.0002 | 0.0015 | 0.0011 | 0.0014 |
| | S | -0.0004 | 0.0031 | -0.0009 | 0.0033 | 0.0011 | 0.0030 |
| | L | -0.0012 | 0.0042 | -0.0006 | 0.0039 | -0.0007 | 0.0038 |
| 2000 | M | 0.0009 | 0.0030 | 0.0013 | 0.0029 | 0.0014 | 0.0029 |
| | S | 0.0002 | 0.0021 | 0.0004 | 0.0022 | 0.0007 | 0.0021 |

Item Difficulty Parameter Recovery

Estimation biases, SEs, and RMSEs of the item difficulty parameter were summarized in Table 3. The results of an analysis of variance with Bias as the dependent variable indicated that the main effect of sample size was statistically significant, $F(2,783) = 3.711$, $p = 0.025$, $f = 0.095$. Tukey *post hoc* tests revealed that estimation biases of item difficulty with SS = 500 was significantly greater than those with SS = 1000 ($p = 0.027$) but not statistically different from those with SS = 2000 ($p = 0.101$). There was also no significant difference in estimation biases between SS = 1000 and SS = 2000 ($p = 0.858$). The main effect of estimation method was not statistically significant, $F(2,783) = 0.168$, $p = 0.845$; neither was the main effect of magnitude of testlet variance, $F(2,783) = 0.227$, $p = 0.797$.

The results of an analysis of variance with SE as the dependent variable indicated that the main effect of sample size was statistically significant, $F(2,783) = 84.200$, $p < 0.001$, $f = 0.464$. Tukey *post hoc* tests revealed that SEs of item difficulty with SS = 500 were significantly greater than those with SS=1000 ($p < 0.001$) and SS = 2000 ($p < 0.001$); SEs of item difficulty with SS = 1000 were also significantly greater than those with SS =2000 ($p < 0.001$). The main effect of estimation method was not statistically significant, $F(2,783) = 0.001$, $p = 0.999$; nor was the main effect of magnitude of testlet variance, $F(2,783) = 0.348$, $p = 0.706$.

The results of an analysis of variance with RMSE as the dependent variable indicated that the main effect of sample size was statistically significant, $F(2,783) = 76.501$, $p < 0.001$, $f = 0.441$. Tukey *post hoc* tests revealed that RMSEs of item difficulty with SS = 500 were significantly greater than those with SS=1000 ($p < 0.001$) and SS = 2000 ($p < 0.001$); RMSEs of item difficulty with SS = 1000 were also significantly greater than those with SS =2000 ($p < 0.001$). The main effect of estimation method was not statistically significant, $F(2,783) = 0.007$, $p = 0.993$; nor was the main effect of magnitude of testlet variance, $F(2,783) = 1.429$, $p = 0.240$.

Table 3

*Parameter Recover for Item Difficulty*

| Estimator | SS | TV | Bias | | SE | | RMSE | |
|---|---|---|---|---|---|---|---|---|
| | | | Mean | SD | Mean | SD | Mean | SD |
| Bayes | 500 | L | 0.0213 | 0.0847 | 0.0490 | 0.0536 | 0.0564 | 0.0611 |
| | | M | 0.0061 | 0.0374 | 0.0421 | 0.0438 | 0.0435 | 0.0447 |
| | | S | 0.0107 | 0.0464 | 0.0455 | 0.0453 | 0.0477 | 0.0477 |
| | 1000 | L | -0.0057 | 0.0571 | 0.0181 | 0.0140 | 0.0212 | 0.0172 |
| | | M | 0.0042 | 0.0383 | 0.0204 | 0.0270 | 0.0218 | 0.0299 |
| | | S | 0.0074 | 0.0499 | 0.0254 | 0.0404 | 0.0278 | 0.0500 |
| | 2000 | L | 0.0008 | 0.0555 | 0.0090 | 0.0085 | 0.0120 | 0.0138 |
| | | M | 0.0037 | 0.0230 | 0.0095 | 0.0103 | 0.0100 | 0.0108 |
| | | S | 0.0049 | 0.0282 | 0.0091 | 0.0084 | 0.0099 | 0.0096 |
| MLR | 500 | L | 0.0215 | 0.0896 | 0.0489 | 0.0536 | 0.0572 | 0.0617 |
| | | M | 0.0063 | 0.0423 | 0.0421 | 0.0436 | 0.0438 | 0.0451 |
| | | S | 0.0118 | 0.0514 | 0.0453 | 0.0458 | 0.0480 | 0.0490 |
| | 1000 | L | -0.0069 | 0.0593 | 0.0180 | 0.0140 | 0.0215 | 0.0175 |
| | | M | 0.0033 | 0.0378 | 0.0203 | 0.0269 | 0.0217 | 0.0298 |
| | | S | 0.0060 | 0.0504 | 0.0251 | 0.0396 | 0.0276 | 0.0488 |
| | 2000 | L | 0.0010 | 0.0578 | 0.0090 | 0.0086 | 0.0123 | 0.0143 |
| | | M | 0.0038 | 0.0253 | 0.0095 | 0.0103 | 0.0102 | 0.0110 |
| | | S | 0.0051 | 0.0302 | 0.0091 | 0.0085 | 0.0100 | 0.0098 |
| WLSMV | 500 | L | 0.0216 | 0.0832 | 0.0525 | 0.0628 | 0.0596 | 0.0691 |
| | | M | 0.0009 | 0.0345 | 0.0419 | 0.0412 | 0.0430 | 0.0418 |
| | | S | 0.0059 | 0.0428 | 0.0427 | 0.0441 | 0.0445 | 0.0464 |
| | 1000 | L | -0.0085 | 0.0562 | 0.0184 | 0.0147 | 0.0215 | 0.0174 |
| | | M | 0.0021 | 0.0355 | 0.0202 | 0.0271 | 0.0214 | 0.0292 |
| | | S | 0.0041 | 0.0454 | 0.0245 | 0.0382 | 0.0265 | 0.0462 |
| | 2000 | L | -0.0004 | 0.0538 | 0.0096 | 0.0101 | 0.0124 | 0.0145 |
| | | M | 0.0034 | 0.0232 | 0.0098 | 0.0110 | 0.0103 | 0.0115 |
| | | S | 0.0040 | 0.0263 | 0.0091 | 0.0082 | 0.0097 | 0.0092 |

*Note.* TV stands for testlet variance, SS for sample size; L stands for Large, M for Medium, S for small.

Item Discrimination Parameter Recovery

Estimation biases, SEs, and RMSES of the item discrimination parameter were summarized in Table 4. The results of an analysis of variance with Bias as the dependent variable indicated that the main effect of sample size was statistically significant, $F(2,783) = 7.233$, $p = 0.001$, $f = 0.135$. Tukey *post hoc* tests revealed that estimation biases of item discrimination with SS = 2000 were significantly smaller than those with SS = 1000 ($p = 0.002$) and SS = 500 ($p = 0.005$). There was no significant difference in estimation biases between SS = 1000 and SS = 500 ($p = 0.962$). The main effect of estimation method was not statistically significant, $F_{(2,783)} = 1.859$, $p = 0.156$; nor was the main effect of magnitude of testlet variance, $F_{(2,783)} = 0.522$, $p = 0.593$. The interaction between sample size and magnitude of testlet variance was statistically significant, $F_{(4, 783)} = 4.087$, $p = 0.003$, $f = 0.143$.

The results of an analysis of variance with SE as the dependent variable indicated that the main effect of sample size was statistically significant, $F_{(2,783)} = 492.283$, $p < 0.001$, $f = 1.121$. Tukey *post hoc* tests revealed that SEs of item discrimination estimation with SS = 500 were significantly greater than those with SS=1000 ($p < 0.001$) and SS = 2000 ($p < 0.001$); SEs of item discrimination estimation with SS = 1000 were also significantly greater than those with SS =2000 ($p < 0.001$). The main effect of magnitude of testlet variance was also statistically significant, $F_{(2,783)} = 5.868$, $p = 0.003$, $f = 0.123$. Tukey *post hoc* tests revealed that SEs of item discrimination with large testlet variance were significantly greater than those with small ($p = 0.002$) but not those with medium testlet variance ($p = 0.349$), and there was no statistically significant difference between medium and small testlet variance ($p = 0.108$). The interaction between sample size and magnitude of testlet variance was also statistically significant, $F_{(4, 783)} = 2.806$, $p = 0.025$, $f = 0.119$. The main effect of estimation method was not statistically significant, $F_{(2,783)} = 1.388$, $p = 0.250$.

The results of an analysis of variance with RMSE as the dependent variable indicated that the main effect of sample size was statistically significant, $F_{(2,783)} = 408.849$, $p < 0.001$, $f = 1.022$. Tukey *post hoc* tests revealed that RMSEs of item discrimination estimation with SS = 500 were significantly greater than those with SS=1000 ($p < 0.001$) and SS = 2000 ($p < 0.001$); RMSEs of item discrimination estimation with SS = 1000 were also significantly greater than those with SS =2000 ($p < 0.001$). The main effect of magnitude of testlet variance was also statistically significant, $F(2,783) = 16.085$, $p < 0.001$, $f = 0.201$. Tukey *post hoc* tests revealed that RMSEs of item discrimination with large testlet variance were significantly greater than those with both those with small ($p < 0.001$) and those with medium testlet variance ($p = 0.001$), and there was no statistically significant difference between medium and small testlet variance ($p = 0.116$). The interaction between sample size and magnitude of testlet variance was also statistically significant, $F(4, 783) = 2.958$, $p = 0.019$, $f = 0.123$. The main effect of estimation method was not statistically significant, $F(2,783) = 1.481$, $p = 0.228$.

Table 4

*Parameter Recover for Item Discrimination*

| Estimator | SS | TV | Bias | | SE | | RMSE | |
|---|---|---|---|---|---|---|---|---|
| | | | Mean | SD | Mean | SD | Mean | SD |
| Bayes | 500 | L | 0.0171 | 0.0575 | 0.0249 | 0.0143 | 0.0284 | 0.0161 |
| | | M | 0.0321 | 0.0383 | 0.0236 | 0.0102 | 0.0261 | 0.0106 |
| | | S | 0.0316 | 0.0338 | 0.0211 | 0.0076 | 0.0233 | 0.0083 |
| | 1000 | L | 0.0336 | 0.0564 | 0.0113 | 0.0053 | 0.0155 | 0.0073 |
| | | M | 0.0233 | 0.0278 | 0.0110 | 0.0048 | 0.0123 | 0.0055 |
| | | S | 0.0174 | 0.0285 | 0.0108 | 0.0045 | 0.0119 | 0.0050 |
| | 2000 | L | 0.0086 | 0.0418 | 0.0051 | 0.0018 | 0.0069 | 0.0045 |
| | | M | 0.0161 | 0.0300 | 0.0053 | 0.0019 | 0.0064 | 0.0028 |
| | | S | 0.0163 | 0.0319 | 0.0049 | 0.0019 | 0.0061 | 0.0034 |
| MLR | 500 | L | 0.0080 | 0.0567 | 0.0229 | 0.0121 | 0.0261 | 0.0142 |
| | | M | 0.0221 | 0.0405 | 0.0219 | 0.0082 | 0.0239 | 0.0091 |
| | | S | 0.0230 | 0.0362 | 0.0196 | 0.0063 | 0.0214 | 0.0075 |
| | 1000 | L | 0.0303 | 0.0577 | 0.0110 | 0.0050 | 0.0151 | 0.0070 |
| | | M | 0.0198 | 0.0282 | 0.0107 | 0.0045 | 0.0119 | 0.0053 |
| | | S | 0.0139 | 0.0298 | 0.0105 | 0.0043 | 0.0116 | 0.0051 |
| | 2000 | L | 0.0053 | 0.0421 | 0.0051 | 0.0017 | 0.0068 | 0.0046 |
| | | M | 0.0131 | 0.0301 | 0.0052 | 0.0019 | 0.0063 | 0.0029 |
| | | S | 0.0134 | 0.0321 | 0.0048 | 0.0019 | 0.0060 | 0.0035 |
| WLSMV | 500 | L | 0.0184 | 0.0572 | 0.0264 | 0.0132 | 0.0299 | 0.0148 |
| | | M | 0.0308 | 0.0389 | 0.0230 | 0.0081 | 0.0255 | 0.0086 |
| | | S | 0.0300 | 0.0307 | 0.0198 | 0.0077 | 0.0216 | 0.0082 |
| | 1000 | L | 0.0375 | 0.0584 | 0.0122 | 0.0056 | 0.0169 | 0.0078 |
| | | M | 0.0259 | 0.0269 | 0.0112 | 0.0047 | 0.0126 | 0.0054 |
| | | S | 0.0195 | 0.0279 | 0.0109 | 0.0043 | 0.0120 | 0.0045 |
| | 2000 | L | 0.0092 | 0.0410 | 0.0057 | 0.0020 | 0.0074 | 0.0042 |
| | | M | 0.0164 | 0.0302 | 0.0056 | 0.0020 | 0.0067 | 0.0028 |
| | | S | 0.0167 | 0.0304 | 0.0049 | 0.0019 | 0.0061 | 0.0031 |

*Note.* TV stands for testlet variance, SS for sample size; L stands for Large, M for Medium, S for small.

## A Real Data Example

In this section we use a real data set to demonstrate the estimation of the 2PL testlet model with the three different methods. The data was extracted from student responses to a test form of the Verbal Section of the General Aptitude Test (GAT-V), a high-stakes test used for college admission purpose in Saudi Arabia. Each GAT-V test has 52 multiple-choice items and 19 or 20 of those are reading comprehension items nested within four passages. In the current

GAT-V test form, there are 19 reading comprehension items, with the third passage having four items, and the other 15 items evenly distributed across the three remaining passages. We drew a random sample of 2,000 students who took the current test form and extracted their responses to the 19 reading comprehension items. The subsequent analyses were based on this 2,000 by 19 response matrix of zeros and ones.

We fit the 2PL testlet model to this dataset using Bayes, MLR, and WLSMV estimators (for Mplus syntax on how to estimate the 2PL testlet model with different estimation methods, see, for example, Luo, 2017). It should be noted that for the BAYES estimator that implements the MCMC algorithm, we specified Mplus to run four parallel chains with each one containing a minimum of 20,000 iterations. The iteration history indicated that the largest PSRF value dropped below 1.1 after 9,000 iterations and at 20,000 iterations, the largest PSRF value was just 1.027, suggesting that model convergence was not an issue. In addition, the Bayesian posterior predictive p (PPP) value is 0.173, indicating an excellent data fit of the 2PL testlet model. In terms of computation time, WLSMV took one second, MCMC 313 seconds, and MLR 322 seconds on a desktop computer with an eight-core Xeon 2.4 GHz processor.

Table 5 lists the testlet variance estimates for the four passages based on the three estimation methods. As can be seen, all methods indicate that the first two passages have small testlet variance, the third passage a negligible testlet variance, and the fourth passage a moderate testlet variance.

Table 5

*Testlet Variance Estimates for the Real Data*

| Estimator | Passage 1 | Passage 2 | Passage 3 | Passage 4 |
|---|---|---|---|---|
| WLSMV | 0.089 | 0.166 | 0.033 | 0.635 |
| MLR | 0.095 | 0.176 | 0.036 | 0.518 |
| BAYES | 0.104 | 0.181 | 0.058 | 0.557 |

Table 6 lists the factor loading and item threshold estimates from the three estimation methods and their converted item parameters based on equations 3-6. It is worth noting again that for WLSMV, only equations 3 and 4 were used for the conversion; for MLR and Bayes, equations 5 and 6 were used first to transform the factor loading and item threshold estimates to be on the same scale of those from WLSMV (which is why the estimates from both MLR and Bayes look very different than those from WLSMV in the table), then equations 3 and 4 were used.

As can be seen, the three sets of converted item parameters are highly similar. We concluded that regardless of the choice of estimation method, the testlet variance estimates and the item difficulty and discrimination parameter estimates for the current dataset are virtually the same, corroborating the findings in the previous section that there are no statistically significant differences regarding parameter recovery between the three estimation methods.

Table 6

*Item Parameter Estimates for the Real Data*

| Item | WLSMV | | | | MLR | | | | BAYES | | | |
|------|-------|-------|------|-------|-------|-------|------|-------|-------|-------|------|-------|
| | $\lambda$ | $\tau$ | $a$ | $b$ | $\lambda$ | $\tau$ | $a$ | $b$ | $\lambda$ | $\tau$ | $a$ | $b$ |
| 1 | 0.73 | -0.53 | 1.92 | -0.73 | 1.14 | -0.83 | 1.93 | -0.73 | 1.15 | -0.84 | 1.94 | -0.73 |
| 2 | 0.76 | -1.31 | 2.12 | -1.72 | 1.29 | -2.21 | 2.21 | -1.70 | 1.30 | -2.22 | 2.23 | -1.70 |
| 3 | 0.42 | -0.09 | 0.80 | -0.21 | 0.47 | -0.10 | 0.80 | -0.21 | 0.47 | -0.10 | 0.80 | -0.21 |
| 4 | 0.53 | -0.07 | 1.08 | -0.13 | 0.63 | -0.09 | 1.08 | -0.15 | 0.63 | -0.09 | 1.09 | -0.15 |
| 5 | 0.51 | 0.27 | 1.03 | 0.53 | 0.61 | 0.32 | 1.05 | 0.52 | 0.61 | 0.32 | 1.06 | 0.52 |
| 6 | 0.60 | -0.37 | 1.34 | -0.62 | 0.78 | -0.49 | 1.31 | -0.63 | 0.78 | -0.49 | 1.31 | -0.63 |
| 7 | 0.53 | 0.10 | 1.10 | 0.19 | 0.61 | 0.12 | 1.04 | 0.20 | 0.61 | 0.12 | 1.04 | 0.20 |
| 8 | 0.64 | -0.09 | 1.51 | -0.14 | 0.86 | -0.12 | 1.47 | -0.14 | 0.87 | -0.12 | 1.47 | -0.14 |
| 9 | 0.54 | -0.11 | 1.13 | -0.20 | 0.66 | -0.15 | 1.13 | -0.22 | 0.66 | -0.15 | 1.14 | -0.22 |
| 10 | 0.21 | 0.28 | 0.37 | 1.33 | 0.21 | 0.29 | 0.35 | 1.40 | 0.21 | 0.29 | 0.35 | 1.40 |
| 11 | 0.60 | -0.76 | 1.29 | -1.27 | 0.72 | -0.94 | 1.22 | -1.31 | 0.72 | -0.94 | 1.23 | -1.31 |
| 12 | 0.75 | -0.62 | 1.97 | -0.83 | 1.15 | -0.96 | 1.98 | -0.84 | 1.16 | -0.97 | 1.94 | -0.85 |
| 13 | 0.13 | 0.68 | 0.22 | 5.23 | 0.15 | 0.69 | 0.26 | 4.53 | 0.15 | 0.69 | 0.26 | 4.53 |
| 14 | 0.10 | 0.23 | 0.17 | 2.30 | 0.10 | 0.23 | 0.17 | 2.30 | 0.10 | 0.23 | 0.17 | 2.30 |
| 15 | 0.60 | -0.20 | 1.29 | -0.33 | 0.77 | -0.26 | 1.29 | -0.35 | 0.77 | -0.26 | 1.30 | -0.35 |
| 16 | 0.30 | 0.27 | 0.54 | 0.90 | 0.30 | 0.29 | 0.52 | 0.97 | 0.30 | 0.29 | 0.50 | 1.00 |
| 17 | 0.38 | 0.53 | 0.74 | 1.39 | 0.42 | 0.60 | 0.73 | 1.39 | 0.42 | 0.60 | 0.71 | 1.43 |
| 18 | 0.26 | 0.54 | 0.47 | 2.08 | 0.29 | 0.57 | 0.49 | 2.00 | 0.29 | 0.57 | 0.49 | 2.00 |
| 19 | 0.31 | 0.33 | 0.57 | 1.06 | 0.34 | 0.36 | 0.57 | 1.06 | 0.34 | 0.36 | 0.57 | 1.06 |

## Conclusions and Discussions

The testlet model is commonly used by researchers and practitioners to address the issue of LID. In contrast to its popularity, psychometric literature on comparison of different estimation methods is scarce for the 2PL testlet model, a popular member of the testlet model family. The current study was intended to fill the gap in literature by comparing the performances of MCMC, MMLE, and WLSMV as estimation methods for the 2PL testlet model.

Utilizing the large number of estimation methods available in Mplus, we implemented the three estimation methods within Mplus and eliminated the potential confounding effect caused by use of different software programs. The results indicated that the estimation method did not significantly affect Bias, SE, and RMSE of testlet variance estimation, or those of item discrimination and difficulty parameter estimation. In other words, the three estimation methods performed similarly regarding estimation of the testlet variance parameter, the item discrimination parameter, and the item difficulty parameter in the 2PL testlet model. Sample size had significant effect on the bias of the testlet variance parameter estimation. Increasing the sample size from 500 to 1000 resulted in significantly smaller bias, but increasing the sample size from 1000 to 2000 did not reduce the bias significantly. Regarding the estimation of item difficulty and discrimination parameters, sample size significantly affected Bias, SE, and RMSE of their estimation. Increasing the sample size from both 500 to 1000 and 1000 to 2000 generally seemed to help improve the estimation quality of both difficulty and discrimination. The magnitude of testlet variance had significant effect on estimation bias of the testlet variance parameter, and SE and RMSE of the item discrimination parameter estimation. For estimation of the testlet variance parameter, decreasing its magnitude from large to medium did not reduce the bias significantly, but decreasing its magnitude from medium to small did lead to significantly smaller bias. For SE of the item discrimination parameter estimation, large testlet variance resulted in significantly smaller SE than small testlet variance; for RMSE of the item discrimination parameter estimation, large testlet variance led to significantly smaller RMSE than both small and medium testlet variance.

The similar performances regarding parameter recovery notwithstanding, the three estimation methods did differ in terms of model convergence rate. MCMC had no issues with model convergence regardless of the sample size and magnitude of testlet variance. When both the sample size and the magnitude of testlet variance were small (SS = 500, TV = 0.25), both MMLE and WLSMV encounter model non-convergence issues, and WLSMV had higher probabilities than MMLE to encounter non-convergence; with the increase of either the sample size or the magnitude of testlet variances, the number of times model non-convergence occurred reduced for WLSMV and disappeared for MMLE.

Another noticeable difference between the three estimation methods was the computation time. MCMC algorithm such as Gibbs sampler and Metropolis Hasting algorithm often require long computation time due to their random walk behavior, especially for complex models such as members in the testlet model family: for example, Jiao et al. (2013) reported that it took WinBUGS approximately 6 hours to fit a 1PL testlet model to a dataset of 1,000 examinees and 54 items. Hamiltonian Monte Carlo (HMC; Neal, 2010) algorithm, a more recent MCMC method, avoids such random walk behavior and reduces the computation time considerably: Luo and Jiao (2017) demonstrated that with the new Bayesian modeling software program Stan (Carpenter et al., 2016), its implementation of the no-U-turn sampler (NUTS; Hoffman & Gelman, 2014) as a variant of HMC, can estimate complex IRT models such as the testlet model and the multilevel IRT model quickly. As the Bayes estimator in Mplus still implements the Gibbs sampler but not the HMC algorithm, we had expected that it would require a substantial amount of computation time. However, as indicated by the average computation time comparison (the simulation study was conducted on a desktop computer equipped with an eight-core 2.40 GHz Xeon processor) between MLR and BAYES as summarized in Figure 1, MLR took considerably more time than BAYES did across all nine simulation conditions. WLSMV was not included here, as its estimation time ranged from one to two seconds.

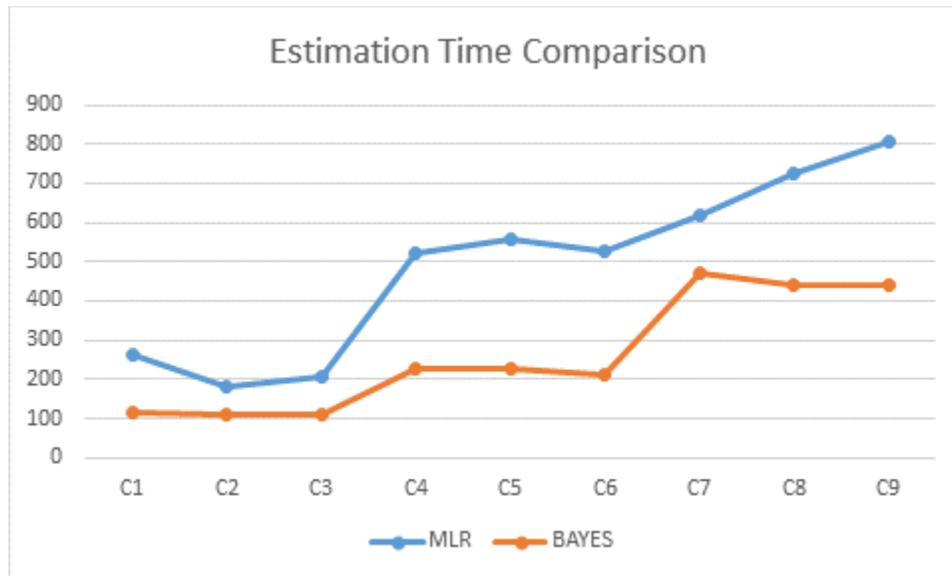

*Figure 1.* Computation time (in seconds) comparison between MLR and BAYES estimators in Mplus across nine simulation conditions

If model convergence is not an issue, WLSMV seems a promising estimation method considering its comparable performances with MMLE and MCMC and its computational speed. As mentioned previously, for the simulated datasets under the current simulation conditions, it took up to two seconds for WLSMV to estimate the 2PL testlet model. It is worth noting that the computation time of WLSMV is known to be sensitive to the number of items other than the number of observations, and it is expected that for a much long test (with 60 items or more), the computation time will be considerably longer to the extent that the speed advantage of WLSMV might totally disappear. However, for tests of medium length, WLSMV is the fastest among the three. In addition, one potential benefit of using limited-information estimation methods for IRT estimation, as pointed out by McDonald and Mok (1995), is that assessment of model fit with can be considerably easier than that with full-information methods.

If model convergence is a concern with small sample size and small magnitude of testlet effect, the MCMC method seems to be the ideal candidate. With the Bayes estimator that implements the MCMC method, Mplus automatically computes the Bayesian PPP value, which can be used for model check purposes; in contrast, the MMLE method provides information criterion based indices such as AIC and BIC, which can only be used for selection of the optimal model among a group of candidates but not assessment of model fit. However, one potential drawback of the Bayes estimator in Mplus is that unlike the WLSMV and MLR estimators, the latent ability estimate (or factor score) cannot be requested directly. One can circumvent this limitation by requesting Mplus to draw multiple plausible values from the posterior distribution of each factor score and obtain the corresponding point estimates by computing the posterior mean (EAP) or the posterior mode (MAP). However, to approximate the posterior distribution accurately, a large number of plausible values may be needed and speed advantage of MCMC over MLR in Mplus might vanish due to the extra computation time required to draw plausible values.

The findings of the current study have important implications for researchers and practitioners. First, the simulation results showed that all three estimation methods, namely MCMC, MMLE, and WLSMV, can be viable estimation methods for the 2PL testlet model, and comparable results can be gained regardless of the choice of estimation method. Second, the current study demonstrated that Mplus, with its provision of various estimation methods and easy-to-learn syntax, is a competitive candidate when it comes to the choice of software for the 2PL testlet model estimation. For those who prefer to use MCMC for the estimation but are concerned with the slow computation of MCMC in Bayesian software such as WinBUGS and its potential non-convergence issue, the Gibbs sampler implemented in Mplus turns out to be quicker than MMLE; in addition, Mplus automatically computes the Bayesian PPP value that can be conveniently used for model check, an attractive feature which comes in handy especially for applied researchers who are not familiar with the WinBUGS syntax enough to program such posterior predictive check procedures themselves.